\documentclass[]{jfm}

\usepackage{graphicx}
\usepackage{newtxtext}
\usepackage{newtxmath}
\usepackage{natbib}
\usepackage{hyperref}
\usepackage{color}
\hypersetup{
    colorlinks = true,
    urlcolor   = red,
    citecolor  = blue,
}

\newcommand{\RomanNumeralCaps}[1]
\linenumbers

\shorttitle{Neutrally buoyant and equally viscous droplets in TC turbulence}
\shortauthor{J Su, Y Zhang, C Wang, L Yi, F Xu, Y Fan, J Wang, and C Sun}

\title{How interfacial tension enhances drag in turbulent Taylor--Couette flow with neutrally buoyant and equally viscous droplets} 

\author{Jinghong Su$^{1}$\footnotemark{},
  Yi-Bao Zhang$^{1}$\footnotemark[\value{footnote}]\footnotetext{Author contributions: J.S. and Y.B.Z. contributed equally to this work.},
  Cheng Wang$^{1}$,
  Lei Yi$^{1}$,
  Fan Xu$^{4}$,
  Yaning Fan$^{1}$,
  Junwu Wang$^{2,4}$
  \corresp{\email{jwwang@cup.edu.cn}},
  Chao Sun$^{1,3}$\corresp{\email{chaosun@tsinghua.edu.cn}}}

\affiliation{\aff{1}New Cornerstone Science Laboratory, Center for Combustion Energy, Key Laboratory for Thermal Science and Power Engineering of Ministry of Education, Department of Energy and Power Engineering, Tsinghua University, 100084 Beijing, China
\aff{2}Beijing Key Laboratory of Process Fluid Filtration and Separation, College of Mechanical and Transportation Engineering, China University of Petroleum, Beijing 102249, P R. China
\aff{3}Department of Engineering Mechanics, School of Aerospace Engineering, Tsinghua University, Beijing 100084, P. R. China
\aff{4}State Key Laboratory of Mesoscience and Engineering, Institute of Process Engineering, Chinese Academy of Sciences, P. O. Box 353, Beijing 100190, P. R. China}
\begin{document}

\maketitle

\begin{abstract}
The presence of dispersed-phase droplets can result in a notable increase in the system's drag. However, our understanding of the mechanism underlying this phenomenon remains limited. In this study, we use three-dimensional direct numerical simulations with a modified multi-marker volume-of-fluid method to investigate liquid-liquid two-phase turbulence in a Taylor--Couette geometry. The dispersed phase has the same density and viscosity as the continuous phase. The Reynolds number $Re\equiv r_i\omega_i d/\nu$ is fixed at 5200, the volume fraction of the dispersed phase is up to $40\%$, and the Weber number $We\equiv \rho u^2_\tau d/\sigma$ is around 8. It is found that the increase in the system's drag originates from the contribution of interfacial tension. 
Specifically, droplets experience significant deformation and stretching in the streamwise direction due to shear near the inner cylinder. Consequently, the rear end of the droplets lags behind the fore head. 
This causes opposing interfacial tension effects on the fore head and rear end of the droplets. For the fore head of the droplets, the effect of interfacial tension appears to act against the flow direction. For the rear end, the effect appears to act in the flow direction. The increase in the system’s drag is primarily attributed to the effect of interfacial tension on the fore head of the droplets which leads to the hindering effect of the droplets on the surrounding continuous phase.
This hindering effect disrupts the formation of high-speed streaks, favoring the formation of low-speed ones, which are generally associated with higher viscous stress and drag of the system. This study provides new insights into the mechanism of drag enhancement reported in our previous experiments.

\end{abstract}


\section{Introduction}

Two-phase flows composed of two immiscible and incompressible liquids are ubiquitous in fields such as petroleum, food, pharmaceuticals, and cosmetics~\citep{spernath2006microemulsions, wang2007oil, mcclements2007critical, kilpatrick2012water}. The last few years have witnessed a renewed interest in two-phase turbulence~\citep{rosti2018rheology,mathai2020bubbly,rosti2021shear, wang2022turbulence,yi2023recent, ni2024deformation,Girotto_Scagliarini_Benzi_Toschi_2024}. The presence of a droplet phase can significantly alter the momentum transport within the flow field, leading to changes in the system's drag. However, the underlying mechanism of droplet-induced drag modulation has not been clearly understood and deserves further investigation, especially in the turbulent regime.

For a liquid-liquid two-phase system, the presence of the droplet phase primarily introduces the influence of the two-phase interface. Consequently, existing studies have focused on examining the effect of the two-phase interface on the system's drag, with particular attention to interfacial dynamics, including deformation, breakup, and coalescence, in shear turbulence.
Interface-resolved direct numerical simulations were employed to study the effect of large deformable droplets on drag enhancement in turbulent channel flow using the interface-capturing method~\citep{scarbolo2015wall,scarbolo2016turbulence}. These studies found that droplet deformability is crucial for droplet-induced drag enhancement; specifically, the stronger the droplet deformability, the weaker the drag enhancement effect.
Interface-capturing simulations of two-phase flow in the Stokes regime were conducted to investigate the effect of coalescence by introducing a short-range repulsive force to prevent droplet merging~\citep{de2019effect}. By comparing the cases allowing droplets to coalesce numerically with the cases using the repulsive force to prevent droplet merging, it was found that droplet coalescence effectively decreases the interfacial surface area, thereby weakening drag enhancement. Conversely, droplet breakup results in higher drag within the system. Interface-capturing simulations of homogeneous isotropic turbulence have also garnered widespread attention for studying the interface's effect on turbulence~\citep{dodd2016interaction, maxey2017droplets, mukherjee2019droplet}.
~\cite{dodd2016interaction} investigated how droplet deformation, breakup, and coalescence affect the temporal evolution of turbulent kinetic energy. They showed that droplet coalescence reduces the total interfacial surface area, causing a decrease in surface energy and an increase in local kinetic energy. Recently, ~\cite{rosti2019droplets} demonstrated that a statistically stationary state (i.e., a balance between coalescence and breakup rates, and convergence of energy balance) can be reached in homogeneous shear turbulence. 
The statistically-stationary state and the balance between coalescence and breakup rates have also been numerically observed in homogeneous isotropic turbulence \citep{mukherjee2019droplet} and wall-bounded turbulence \citep{soligo2019breakage}.
Therefore, from the perspective of interfacial dynamics, droplet deformation plays a key role in drag enhancement.

Related experiments have also investigated drag modulation in liquid-liquid two-phase turbulence~\citep{bakhuis2021catastrophic, yi2021global, wang2022turbulence, yi2022physical}. In Taylor--Couette turbulence, it has been experimentally observed that when breakup events dominate at low droplet volume fractions, dispersed droplets exhibit a specific size distribution well-described by the lognormal distribution~\citep{yi2021global}. At a fixed Reynolds number, droplets maintain nearly the same average size, while the system's drag increases with the droplet volume fraction~\citep{yi2022physical}. The effective viscosity as a function of volume fraction shows an increasing trend with the dispersed phase volume fraction, similar to that observed in rigid particle suspensions at moderate volume fractions ($\phi=0-40\%$) \citep{krieger1959mechanism}, although the effective viscosity for rigid particles consistently exceeds that for dispersed droplets \citep{yi2021global}. Numerical results from~\cite{de2019effect} have recently reported a similar phenomenon. However, the underlying mechanisms remain elusive.

Given the challenge of obtaining sufficient experimental data, interface-resolved simulations of liquid-liquid Taylor--Couette flow at a Reynolds number $Re=960$ were conducted by~\cite{hori2023interfacial}. They reported that the system can be categorized into two regimes based on the Weber number $We$: advection-dominated and interface-dominated regimes for high and low $We$, respectively. In the advection-dominated regime, drag enhancement as a function of volume fraction shows a non-monotonic behavior. This could be partly attributed to the lower Reynolds number in their study compared to experimental conditions \citep{yi2021global}, and the significant numerical challenges related to coalescence encountered with the adapted interface-capturing method used for the volume fractions studied~\citep{elghobashi2019direct}.
Recent works~\citep{su2024turbulence,su2024numerical} numerically investigated the effect of drops with varying density and viscosity in turbulent Taylor--Couette flow. It was found that interfacial tension consistently enhances momentum transport, thereby contributing to drag enhancement. However, the specific mechanism by which interfacial tension induces drag enhancement remains unclear.

In the interface-capturing method, merging of interfaces occurs automatically whenever two interfaces come within one grid cell of each other~\citep{elghobashi2019direct}. This makes it very difficult to simulate droplets with moderate or high volume fractions. Various methods have been employed to address this issue, such as adaptive grid refinement~\citep{innocenti2021direct}, film drainage models~\citep{thomas2010multiscale}, artificial repulsive forces~\citep{de2019effect}, and multi-marker methods~\citep{coyajee2009numerical}. However, these methods are primarily effective when dealing with a small number of droplets. Simulating two-phase turbulence with hundreds or more droplets becomes prohibitively expensive due to significantly increased computational costs. In this work, we investigate liquid-liquid Taylor--Couette turbulence at moderate volume fractions ($\phi=0-40\%$) using a modified multi-marker volume-of-fluid method. This modified approach allows us to reproduce the experimental drag modulation results reported in~\cite{yi2021global}. Our objective is to gain an intuitive understanding of how the droplet phase modulates the system's drag.

The manuscript is organized as follows: In \S 2, we describe the numerical method and setup. In \S 3, we discuss the effect of droplets on the angular velocity flux and analyze the modulation mechanism of viscous stress and angular velocity within the boundary layer. Finally, conclusions are drawn in \S 4.


\section{Numerical method and setting}\label{sec2}
The droplet and continuous phases are considered to be immiscible, incompressible and Newtonian.
The two-phase immiscible and incompressible flow is governed by the Navier-Stokes equations 
\begin{equation}
  \nabla\cdot\boldsymbol{u}
  =0,
\end{equation}
\begin{equation}
  {\partial _t(\rho \boldsymbol{u})}
  +\nabla\cdot(\rho\boldsymbol{uu})
  =
  -{\nabla{p}}
  +\nabla\cdot{\boldsymbol{\tau}}
  +\boldsymbol{f_\sigma},
\end{equation}
where $\boldsymbol{u}$ is the velocity field, $p$ is the pressure, $\boldsymbol{\tau} = \mu(\nabla\boldsymbol{u} + (\nabla{\boldsymbol{u}})^T)$ is the viscous stress tensor, and $\boldsymbol{f_\sigma}$ represents the interfacial tension. Here, $\rho$ and $\mu$ denote the density and viscosity of the combined phase.
In the standard volume-of-fluid (VOF) method, merging of interfaces occurs automatically whenever two interfaces come within one grid cell of each other, a phenomenon known as numerical coalescence~\citep{elghobashi2019direct,soligo2021turbulent}. This makes it challenging to simulate droplets with moderate to high volume fractions. To address this issue, this work employs a modified multi-marker VOF method which was originally proposed by~\cite{coyajee2009numerical} to prevent numerical coalescence. Specifically, multiple markers, $\alpha_m$, are introduced to represent each group of droplets, where each droplet group is individually marked. 
The range of $\alpha_m$ is from zero to one: $\alpha_m=0$ represents the continuous phase, $\alpha_m=1$ represents the dispersed phase, and $0<\alpha_m<1$ denotes the interface region. This modified multi-marker VOF method enables the simulation of two-phase flow at moderate droplet volume fractions. The evolution of $\alpha_m$ is governed by the transport equation
\begin{flalign}
 && && {\partial _t \alpha_m}
  +\nabla\cdot(\alpha_m \boldsymbol{u})
  =0,
  && m=1,...,n. && &&
  \vspace{-0mm}
    \label{equ:marker}
    \end{flalign}
$n$ is the total number of the markers. The interfacial tension is calculated by
 \begin{equation}
 \boldsymbol{f_\sigma}=\sum_{m=1}^n\sigma\kappa_m\nabla\alpha_m,
   \vspace{-0mm}
\end{equation}
where $\sigma$ is the surface tension coefficient and $\kappa_m = -\nabla\cdot({\nabla\alpha_m}/{|\nabla\alpha_m|})$ is the interface curvature. The density $\rho$ and viscosity $\mu$ of the combined phase are defined as functions of the phase fraction $\alpha$, specifically $\rho=\alpha\rho_d + (1-\alpha)\rho_f$ and $\mu=\alpha\mu_d + (1-\alpha)\mu_f$, where $\rho$ and $\mu$ with subscripts $f$ and $d$ denote the density and viscosity of the continuous phase and dispersed phase, respectively. The phase fraction $\alpha$ is defined as $\alpha = \max\{\alpha_1, \ldots, \alpha_n\}$~\citep{coyajee2009numerical}. In this work, we use $\rho=\rho_f=\rho_d$ and $\mu=\mu_f=\mu_d$. 

We consider a two-phase flow in Taylor--Couette (TC) turbulence, where the flow is confined between two coaxial cylinders with radii $r_i$ (inner) and $r_o$ (outer). The curvature of the Taylor--Couette system is characterized by the ratio $\eta=r_i/r_o=0.714$. The outer cylinder (OC) is fixed, while the inner cylinder (IC) rotates at a constant angular velocity $\omega_i$.
The torque $T$ required to drive the IC is examined to study the droplet-induced drag enhancement.
To minimize computational costs without compromising accuracy, we employ a rotational symmetry of order 6 (i.e., the azimuthal angle of the simulated domain is $\pi/3$) and an aspect ratio of $\Gamma=L/d=2\pi/3$ in the simulated Taylor--Couette system, where $d$ is the gap width between the cylinders and $L$ represents the axial length. This choice has been validated for both single-phase and multiphase Taylor--Couette turbulence~\citep{brauckmann2013direct, spandan2018physical, assen2022strong}.

In this work, liquid-liquid TC turbulence is simulated with total droplet volume fractions of $\phi=0$, $\phi=10\%$, $\phi=20\%$, $\phi=30\%$, and $\phi=40\%$.  The Reynolds number $Re\equiv{\rho} u_i d/\mu$ is fixed to 5200, where $u_i = r_i \omega_i$ is the velocity of the IC. The Taylor number $Ta \equiv \chi(r_o+r_i)^2(r_o-r_i)^2\omega_i^2/(4\nu^2) = 4.12\times 10^7$, where $\chi = [(r_i+r_o)/(2\sqrt{r_i r_o})]^4$ and $\nu=\mu/\rho$. The frictional Reynolds number at the IC $ Re_\tau\equiv u_\tau d/\nu$ is 264.86, 281.62, 297.01, 315.37 and 329.76 for $\phi=0$, $10\%$, $20\%$, $30\%$ and $40\%$, respectively. $u_\tau$ is the friction velocity and defined as $\sqrt{\tau_{w}/\rho}$, where $\tau_{w}$ represents the shear stress at the IC. The Weber number $We \equiv \rho u_\tau^2 d/\sigma$ is 7.96, 8.74. 9.85 and 10.77 for $\phi=10\%$, $20\%$, $30\%$ and $40\%$, respectively.

No-slip and impermeable boundary conditions are imposed on both cylinder surfaces, while periodic boundary conditions are applied in the axial and azimuthal directions. The inner cylinder (IC) and outer cylinder (OC) are subjected to Neumann boundary conditions for the phase fraction, resulting in a default contact angle of $90^\circ$.
The Taylor--Couette (TC) system is discretized using a collocated grid system consisting of $N_\theta \times N_r \times N_z = 336 \times 250 \times 192$ points in the azimuthal, radial, and axial directions, respectively. The grids are uniformly distributed in the azimuthal and axial directions but are unevenly spaced and concentrated near the two cylinders in the wall-normal direction.
The grid spacing is measured in units of the viscous length scale $\delta_\nu = \nu/u_\tau$ for single-phase flow. In the radial direction, the grid spacing varies from $0.31\delta_\nu$ near the wall to $2.45\delta_\nu$ at the center of the gap. In the azimuthal direction, it varies from $2.06\delta_\nu$ near the IC to $2.89\delta_\nu$ near the OC. The grid spacing is uniform in the axial direction, with a value of $2.89\delta_\nu$.
The Kolmogorov scale, denoted as $\eta_k$, is $2.07\delta_\nu$ for single-phase turbulence, determined by the exact dissipation relationships given by $\eta_k/d=(\chi^{-2}Ta(Nu_{\omega}-1))^{-1/4}$, where $Nu_\omega=J^{\omega}/J^{\omega}_{lam}$~\citep{eckhardt2007torque}. Here, $J^{\omega}$ represents the total angular velocity flux and $J^{\omega}_{lam}$ corresponds to the flux under fully laminar and non-vortical conditions. The maximum grid spacing is $1.39\eta_k$ for single-phase turbulence.
To ensure sufficient spatial resolution to resolve the smallest length scales, a resolution test is performed for the single-phase case (see Appendix~\ref{appA}).

These simulations utilize the modified multi-marker VOF method with a piecewise-linear interface calculation (PLIC) algorithm implemented in the interFoam solver of the open-source OpenFOAM v8~\citep{rusche2003computational, chen2022turbulent}. 
Based on the PLIC algorithm, the resolved interface region ($0<\alpha<1$) could be confined within a single layer of grid cells between the two phases~\citep{su2024numerical}. Therefore, the PLIC algorithm also works in reducing the influence of numerical coalescence.
The robustness of OpenFOAM in simulating both single-phase and two-phase Taylor--Couette (TC) turbulence has been demonstrated in our previous works~\citep{xu2022direct, xu2023direct, su2024numerical}.
The maximum Courant-Friedrichs-Lewy number is set to 0.2. Temporal discretization employs a blended scheme between the first-order Euler scheme and the second-order Crank-Nicolson scheme, with a blending factor of 0.9 for robustness and accuracy. Spatial discretization uses a second-order linear-upwind scheme for the advection term in the momentum equation. The phase fraction transport equation employs a piecewise-linear interface calculation scheme to maintain interface sharpness.
The PIMPLE algorithm~\citep{holzmann2016mathematics}, which is a hybrid version of the PISO algorithm and the SIMPLE algorithm, is used to handle the pressure–velocity coupling to guarantee better stability for two-phase simulations. The pressure equation is solved using the Geometric Algebraic Multigrid solver coupled with the Simplified Diagonal-based Incomplete Cholesky smoother, which is commonly used to speed up the computational efficiency in simulating two-phase flow~\citep{scheufler2019accurate,chen2022turbulent}. Velocity and phase fraction are solved using an iterative solver with a symmetric Gauss-Seidel smoother. In the simulations, a residual tolerance of $10^{-6}$ is maintained for all variables, except for the phase fraction, which has a tolerance of $10^{-8}$.
The computational accuracy of these settings has been verified by comparing our results with those of~\cite{ostilla2013optimal} in our previous work~\citep{su2024turbulence,su2024numerical}. 

For a fixed volume fraction $\phi$ in the modified multi-marker VOF method, 
the dispersed phase is divided into $n$ groups, and each group is assigned a marker $\alpha_m$, and the droplet volume fraction for each marker $\phi_{mvf}$ is the same. A large value of $\phi_{mvf}$ results in unphysical numerical coalescence; while a too small value of $\phi_{mvf}$ limits the maximum droplet size and is computationally costly as one needs to solve $n$ transport equations. In our study, we choose $\phi_{mvf}=5\%$ for all cases. Specifically, $n=2$, 4, 6, and 8 for droplet volume fractions of $10\%$, $20\%$, $30\%$, and $40\%$, respectively. This choice is inspired by previous studies \citep{rosti2019droplets,soligo2019breakage,crialesi-esposito_rosti_chibbaro_brandt_2022,Mangani_Roccon_Zonta_Soldati_2024}, which used the standard volume-of-fluid method or phase-ﬁeld method. In these studies, the volume fraction is mainly in the range of $3 \sim 10 \%$, where the effect of numerical coalescence is considered to be negligible. The choice of $\phi_{mvf}=5\%$ faithfully reproduces the global drag of the system reported in our previous experimental study \citep{yi2021global}. To ensure that the modified multi-marker VOF method does not qualitatively alter the drag modulation mechanism, two additional simulations using the standard VOF method are conducted and compared with the multi-marker VOF method (see Appendix~\ref{appB}). During our simulations, a single-phase case is first simulated to initialize the velocity field. Once a fully-developed flow with a pair of Taylor rolls is obtained, droplets of diameter $0.2d$ are uniformly positioned in the simulated domain $\Omega$. The droplets are then randomly marked by the marker $\alpha_m$ ($m=1,\cdots n$) in such a way that the total volume of the droplets for each marker is $\phi_{mvf}V_\Omega$, where $V_\Omega$ is the volume of the simulated domain.
All statistics presented in the paper are collected over at least $3 \times 10^2$ large eddy turnover times, defined as $(r_o-r_i)/(\omega_i r_i)$, after the two-phase flow reaches a statistically steady state.

\begin{figure}
	\centering
	\includegraphics[width=1\linewidth]{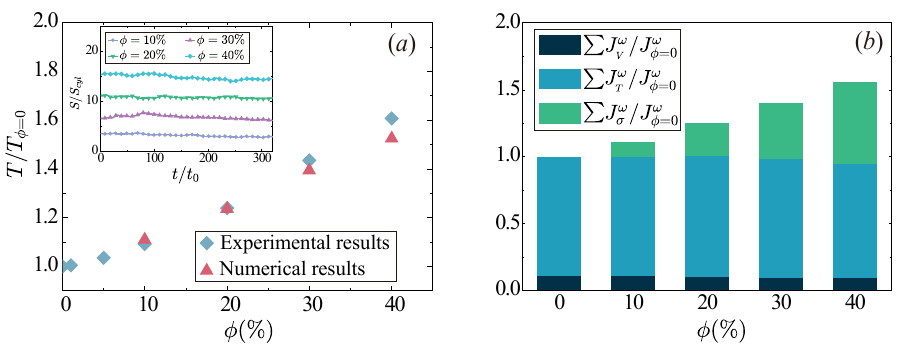}
	\caption{(\textit{a}) The torque $T$ needed to drive the IC rotating at a constant rate $\omega_i$. $T$ is normalized by the single-phase torque $T_{\phi=0}$. The simulated results are compared with experimental results from our previous work~\citep{yi2021global}. The inset shows the interfacial surface area $S/S_{cyl}$ as a function of time $t/t_0$ at various droplet volume fraction $\phi$, where $S_{cyl}$ is the surface area of the IC and $t_0$ is the large eddy turnover time. (\textit{b}) The turbulent stress $J^{\omega}_{_T}$, viscous stress $J^{\omega}_{_V}$, and interfacial stress $J^{\omega}_\sigma$ contribution to the angular velocity flux. The operator $\sum$ denotes radially-averaged quantity. All the contributions are normalized by the total angular velocity flux for the single-phase flow $J_{\phi=0}^{\omega}$.
	}
	\label{fig:1}
\end{figure}

To demonstrate that the flow reaches a statistically steady state, the temporal evolution of the interfacial surface area $S$ is shown in the inset of figure~\ref{fig:1}(\textit{a}). $S$ fluctuates around a constant value dependent on $\phi$, indicating a balance between droplet breakup and coalescence within the system.
In our simulations, the same grid system is employed for both single-phase and two-phase cases, and a resolution test is conducted for both the $\phi=0$ and $\phi=10\%$ scenarios (see Appendix~\ref{appA}). Due to the significant increase in computational cost with the multi-marker VOF method, conducting a resolution test for $\phi=40\%$ becomes impractical. The corresponding grid spacing near the wall is approximately $1.39\eta_k$ for $\phi=0$ and $1.58\eta_k$ for $\phi=40\%$, indicating only mild degradation in spatial resolution with increasing $\phi$.
In figure~\ref{fig:1}(\textit{a}), we present the torque $T$ required to rotate the IC at a constant rate $\omega_i$, superimposed with experimental data from ~\cite{yi2021global} for comparison purpose. It is evident that $T$ increases with $\phi$ in both numerical and experimental data, indicating drag enhancement due to dispersed interfaces. The datasets agree well with each other at $\phi\leq 30\%$, with a minor deviation observed at $\phi=40\%$. However, the method and grid allocation adopted here suffice for the purpose of this study: revealing the underlying mechanism of drag enhancement by dispersed droplets. 

\section{Results and Discussion}
In this work, we fix the OC while sustaining the constant rotational velocity of the IC. The torque $T$ required to drive the IC is examined to study the underlying mechanism of drag enhancement by dispersed droplets. In the TC system, the angular velocity flux $J^{\omega}$ is conserved in the radial direction. $J^\omega$ and the torque at IC $T$ are related through the equation $T=2\pi L J^\omega$~\citep{eckhardt2007torque}, which provides an efficient way to study the mechanism of the drag enhancement. Inspired by an analogous situation in turbulent channel flows~\citep{picano2015turbulent,wang2023numerical}, where it is conventional to decompose the total stress into the viscous part, turbulent part and others, we decompose $J^\omega$ into contributions from the turbulent stress $\tau_{_T}$, the viscous stress $\tau_{_V}$ and the interfacial tension $\boldsymbol{f_\sigma}$~\citep{eckhardt2007torque,su2024numerical}, i.e.,
 \begin{equation}
 J^\omega\equiv J_{_T}^\omega(r) + J_{_V}^\omega(r)+J_{\sigma}^\omega(r)= \rm{const.},
   \label{equ:adv}
 \vspace{0mm}
 \end{equation}
where the three terms represent

(i) the turbulent stress contribution $J_{_T}^\omega(r) =r^3 \langle \rho u^{\prime}_r \omega^{\prime}\rangle = r^2 \langle\tau_{_T}\rangle$,

(ii) the viscous stress contribution $J_{_V}^\omega(r) = -\mu r^3 \partial _r \langle \omega \rangle = r^2 \langle\tau_{_V}\rangle$,

(iii) the interfacial tension contribution $J_{\sigma}^\omega(r) =-\int_{r_i}^{r} \left \langle {r^2 f_\sigma^\theta}\right \rangle  \, {\rm{d}}r$.\\
Here, $u^{\prime}_r=u_r-\langle u_r \rangle$, $\omega^{\prime}=\omega-\langle \omega \rangle$, $r$ is the radial position, $u_r$ is the radial velocity, $\omega$ is the angular velocity and $f_\sigma^\theta$ is the azimuthal component of interfacial tension. The operator $\left \langle \cdot \right \rangle$ denotes the average in the axial and azimuthal directions and over time. In this work, all contributions are normalized by the total angular velocity flux for single-phase flow $J_{\phi=0}^{\omega}$. 
We note that $J_{_T}^\omega(r)$, $J_{_V}^\omega(r)$, and $J_\sigma^\omega(r)$ all depend on the radial position. To eliminate the radial dependence, $J_{_T}^\omega(r)$, $J_{_V}^\omega(r)$, and $J_\sigma^\omega(r)$ are further averaged in the radial direction, denoted as $\sum J_{_T}^\omega$, $\sum J_{_V}^\omega$, and $\sum J_\sigma^\omega$, and are depicted in figure~\ref{fig:1}(\textit{b}). This approach has been widely used to study drag modulation in two-phase flows~\citep{de2019effect,ardekani2019turbulence,hori2023interfacial,wang2023numerical}. Both $\sum J_{_T}^\omega$ and $\sum J_{_V}^\omega$ are virtually not changed by the dispersed droplets. However, $\sum J_\sigma^\omega(r)$ increases monotonically with $\phi$. Figure \ref{fig:1}(\textit{b}) demonstrates that the drag enhancement is dominated by the interfacial tension contribution, consistent with our previous findings~\citep{su2024turbulence,su2024numerical}.

\begin{figure}
	\centering
	\includegraphics[width=0.95\linewidth]{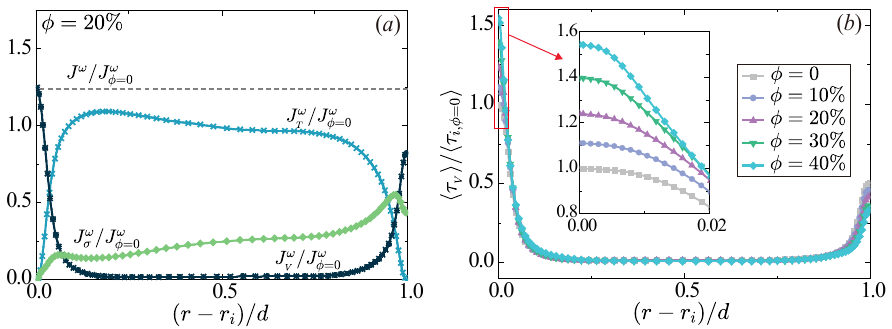}
	\caption{(\textit{a}) Angular velocity flux and its three contributions as a function of the radial position for the case with $\phi=20\%$. The radial position $(r-r_i)/d=0$ corresponds to the IC and $(r-r_i)/d=1.0$ corresponds to the OC. (\textit{b}) Viscous stress $\left \langle \tau_{_V} \right \rangle$ (normalize by the single phase value at $(r-r_i)/d=0$, i.e., $\left\langle \tau_{i,\phi=0} \right\rangle$) as a function of the radial position at various droplet volume fractions. Inset: zoom-in of $\left \langle \tau_{_V} \right \rangle/\left \langle \tau_{i,\phi=0} \right \rangle$ near the IC.}
	\label{fig:2}
\end{figure}

In the above analysis, drag enhancement is attributed to the additional contribution of interfacial tension to the angular velocity flux. However, it remains unclear how the introduction of the interface increases the overall drag of the system. Figure \ref{fig:1}(\textit{b}) gives us the impression that viscous stress plays a negligible role in transporting the angular velocity flux. However, this is not the case, especially in the boundary layer, where viscous stress dominates over turbulent stress. It should be noted that averaging in the radial direction masks the important role of the viscous boundary layer, as it represents only a small fraction of the total volume.
Taking the droplet volume fraction $\phi=20\%$ as an example, $J_{_V}^\omega/J^\omega_{\phi=0}$ is much larger than $J_{_T}^\omega/J^\omega_{\phi=0}$ and $J_\sigma^\omega/J^\omega_{\phi=0}$ near the IC, as shown in figure~\ref{fig:2}(\textit{a}). Due to the presence of a positive interfacial tension contribution near the OC, a smaller viscous stress contribution ($J_{_V}^{\omega}/J_{\phi=0}^{\omega}<1$) is observed, which has also been reported by~\cite{hori2023interfacial}. In our previous study, we observed that droplets are fragmented by dynamic pressure within the boundary layer \citep{yi2022physical}. Given the higher observed viscous stress contribution compared to the single-phase case at the IC surface (figure \ref{fig:2}\textit{a}), further analysis of the boundary layer near the IC may yield insights into the underlying mechanism of drag enhancement by dispersed droplets.

Since the viscous stress contribution $J_{_V}^{\omega}(r)$ for a specific radial position depends solely on the viscous stress $\tau_{_V}$, we will now focus our attention on $\tau_{_V}$ for ease of discussion. We show the $\tau_{_V}$ distribution at various droplet volume fractions in figure~\ref{fig:2}(\textit{b}) and an enlarged view of $\tau_{_V}$ within the region $(r-r_i)/d<0.02$ is displayed in the inset of figure~\ref{fig:2}(\textit{b}). $(r-r_i)/d=0.02$ corresponds to $y^+=5.31$, 5.61, 5.94, 6.31 and 6.60 for cases with $\phi=0$, $10\%$, $20\%$, $30\%$ and $40\%$, respectively. $y^+$ is the distance from the IC in unit of the viscous length scale at the wall. In the boundary layer, $\tau_{_V}$ increases with increasing volume fraction, in line with the drag enhancement observed in figure \ref{fig:1}.

\begin{figure}
	\centering
	\includegraphics[width=0.95\linewidth]{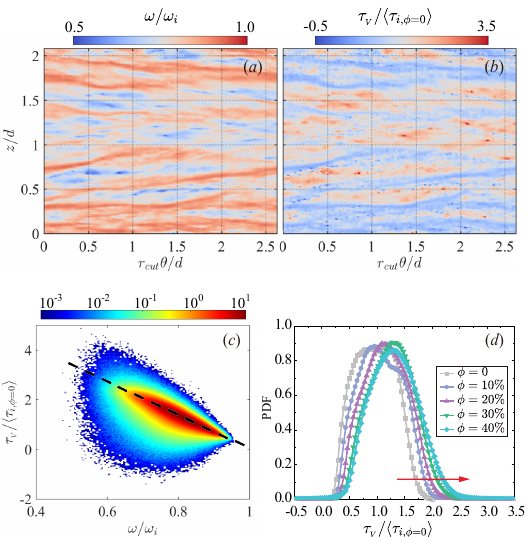}
	\caption{Contour plots of (\textit{a}) the instantaneous angular velocity and (\textit{b}) viscous stress on the cylinder surface with radius $(r_{cut}-r_i)/d=0.0105$ for two-phase turbulence with $\phi=40\%$. $(r_{cut}-r_i)/d=0.0105$ corresponds to $y^+=3.45$. (\textit{c}) Joint probability density function between the angular velocity and the viscous stress for $\phi=40\%$.  (\textit{d}) Probability density functions of $\tau_{_V}$ for different droplet volume fractions. In (\textit{c}) and (\textit{d}), the data are sampled on the same cylinder surface as in (\textit{a}) and (\textit{b}).}
	\label{fig:3}
\end{figure}

Given the definition of viscous stress $\tau_{_V} = -\mu r \partial _r \omega$, the magnitude of viscous stress is determined by the angular velocity gradient when viscosity and $r$ are fixed. Considering that the IC has a constant angular velocity, it is likely that in the boundary layer the viscous stress is directly related to the angular velocity. Figures~\ref{fig:3}(\textit{a}) and~\ref{fig:3}(\textit{b}) show contour plots of the instantaneous angular velocity $\omega$ and viscous stress $\tau_{_V}$ on a cylinder surface with $(r_{cut}-r_i)/d=0.0105$ for the $\phi=40\%$ case. By carefully examining their spatial distribution, it can be observed that the low-speed streak regions (blue color in figure~\ref{fig:3}\textit{a}) predominantly correspond to the high viscous stress regions  (red color in figure~\ref{fig:3}\textit{b}). To provide quantitative evidence for this observation, we calculate the joint probability density function (PDF) between $\tau_{_V} / \left\langle  \tau_{i,\phi=0}\right\rangle $ and $ \omega / \omega_i$. The joint PDF basically lies along the straight dashed line with a negative slope (see figure~\ref{fig:3}\textit{c}), suggesting a negative correlation between $\tau_{_V} / \left\langle  \tau_{i,\phi=0}\right\rangle $ and $ \omega / \omega_i$: namely, lower velocity fluid can result in higher viscous stress. As the volume fraction increases, the PDFs of the viscous stress shift rightward, implying an overall increase in $\tau_{_V}$ (see figure~\ref{fig:3}\textit{d}). Based on the negative correlation between $\tau_{_V} / \left\langle  \tau_{i,\phi=0}\right\rangle $ and $ \omega  / \omega_i$, the rightward shift of $\tau_{_V}$ means that the angular velocity decreases. Here, we reveal the negative correlation between viscous stress and angular velocity within the boundary layer, where an increase in viscous stress manifests itself as a decrease in angular velocity. 
Figure~\ref{fig:3}(\textit{d}) demonstrates that an overall increase in viscous stress is observed with increasing droplet volume fraction, which would correspond to a decrease in angular velocity.
Therefore, understanding the reason for the decrease in angular velocity within the boundary layer becomes crucial for the study of drag enhancement.

\begin{figure}
	\centering
	\includegraphics[width=1\linewidth]{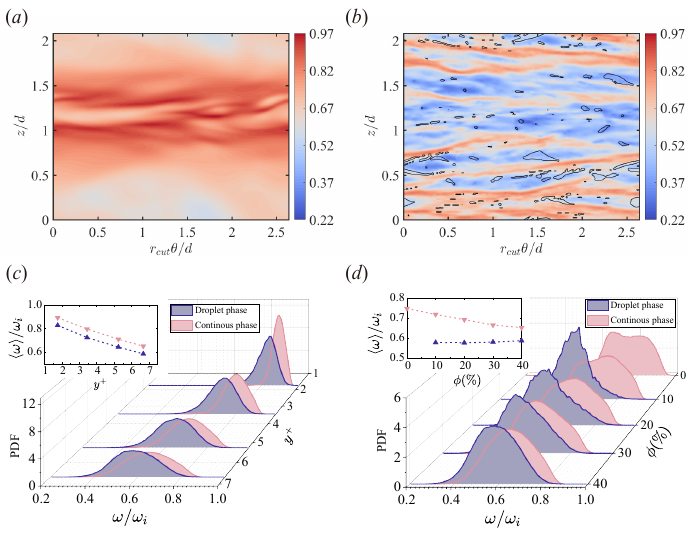}
	\caption{  Contour plots of the instantaneous angular velocity $\omega/\omega_i$ on a cylinder surface with $(r_{cut}-r_i)/d=0.02$ for (\textit{a}) single-phase turbulence with $\phi=0$ and (\textit{b}) two-phase turbulence with $\phi=40\%$. The droplet interfaces ($\alpha=0.5$, solid lines) are also superimposed in (\textit{b}). (\textit{c}) Probability density functions of angular velocity for the droplet and continuous phases at different cylinder surfaces for $\phi=40\%$ case. The inset shows the average angular velocity of the droplet and continuous phases as a function of wall distance. (\textit{d}) Probability density functions of angular velocity for the droplet and continuous phases at fixed cylinder surface $(r_{cut}-r_i)/d=0.02$ for various droplet volume fractions. The inset shows the average angular velocity of the droplet and continuous phases.}
	\label{fig:4}
\end{figure}

To visually compare the angular velocity distribution in single-phase and two-phase turbulence, contour plots of the instantaneous angular velocity for cases with $\phi=0$ and $\phi=40\%$ are illustrated in figures~\ref{fig:4}(\textit{a}) and \ref{fig:4}(\textit{b}), respectively. Herring-bone streaks can be observed in single-phase turbulence due to centrifugal instability~\citep{barcilon1979marginal,dong2007direct}. Compared to single-phase turbulence, two-phase turbulence exhibits numerous low-speed streaks. The high-speed streaks are disrupted by the low-speed ones, and their axial size diminishes. Droplet interfaces ($\alpha=0.5$) are depicted as solid lines in figure \ref{fig:4}(\textit{b}). Upon careful inspection, it is observed that droplets predominantly occupy the low-speed streaks. This observation suggests that droplets tend to favor the formation of low-speed streaks. Subsequently, we demonstrate that the presence of droplets indeed impedes the continuous phase, disrupts the high-speed streaks, and promotes the low-speed ones.

The PDF of angular velocity for both the continuous and droplet phases are shown in figures~\ref{fig:4}(\textit{c}) and \ref{fig:4}(\textit{d}). In figure \ref{fig:4}(\textit{c}), the droplet volume fraction is $\phi=40\%$ while the data are sampled on different cylinder surfaces $(r_{cut}-r_i)/d=0.0054$, 0.0105, 0.0159 and 0.02, which correspond to $y^+=1.77$, 3.45, 5.25 and 6.60. The first two surfaces are located inside of the viscous sublayer and the last two are in the buffer layer. In figure \ref{fig:4}(\textit{d}), the data are sampled on the cylinder surface with $(r_{cut}-r_i)/d=0.02$ while two-phase turbulence with different volume fractions are considered. It is observed that the angular velocity of the droplet phase is consistently lower than that of the continuous phase (see figure \ref{fig:4}\textit{c} and its inset). The observation persists across various volume fractions investigated here (see figure~\ref{fig:4}\textit{d}). Besides, at the fixed radial position $(r_{cut}-r_i)/d=0.02$, the average angular velocity of the droplet phase is nearly invariant at different droplet volume fractions, whereas the average angular velocity of the continuous phase decreases as the droplet volume fraction increases (see inset of figure~\ref{fig:4}\textit{d}). The lower velocity of the droplets than the continuous phase suggests that the droplet phase will impede the continuous phase surrounding the droplets.

\begin{figure}
	\centering
	\includegraphics[width=1\linewidth]{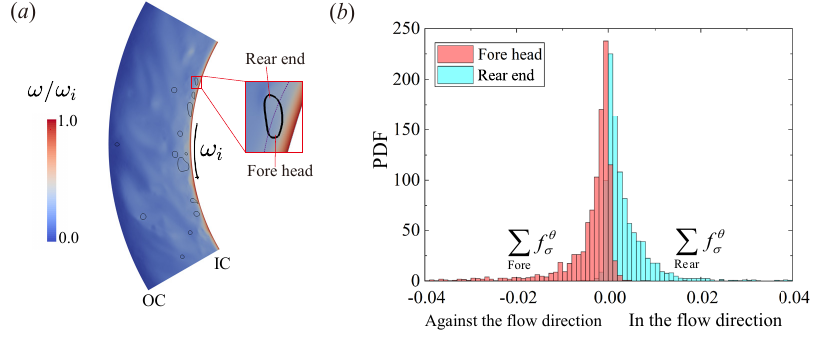}
	\caption{(\textit{a}) Contour plot of the instantaneous angular velocity in the $r-\theta$ plane for $\phi=40\%$ case. Droplet interfaces ($\alpha=0.5$) are represented by the solid lines. The inset is an enlarged view of the main figure marked by the red rectangle. The droplet in the inset is divided into fore head and rear end along the radial direction based on the location of the droplet’s center of mass. (\textit{b}) The PDF of the sum of the azimuthal interfacial tension acting on the fore head $\sum\limits_{\scriptscriptstyle\text{Fore}}{f_{\sigma}^{\theta}}$ and the sum of the azimuthal interfacial tension acting on the rear end $\sum\limits_{\scriptscriptstyle\text{Rear}}{f_{\sigma}^{\theta}}$ are shown for droplets with centers of mass in the range $y^+<24$. The two terms in (\textit{b}) are normalized by the absolute value of the total azimuthal interfacial tension experienced by all droplets with centers of mass in the range $y^+<24$. Here, the total azimuthal interfacial tension is negative, indicating that the overall effect of these droplets is to impede the surrounding flow.}
	\label{fig:5}
\end{figure}

To explain why the droplet phase moves slower than the continuous phase near the IC, we show a cross-section of flow in the $r-\theta$ plane for $\phi=40\%$ case in figure \ref{fig:5}(\textit{a}) and ideally divide a single droplet in the vicinity of the IC into two parts along the radial direction based on the location of the droplet's center of mass.
The part closer to the IC is called fore head of the droplet, and the other part is called the rear end of the droplet.
Due to the high shear near the IC, droplets are strongly deformed. However, the droplets do not align with the azimuthal shear direction but show a slight deviation. This deviation from the $\theta$ direction originates from the fact that fluid parcel closer to the rotating IC will generally have a higher angular velocity. In other words, the fore head of the droplet moves faster than its rear end. To maintain the integrity of the droplet and resist deformation, the rear end will drag the fore head backward due to the effect of interfacial tension. This process will make the fluid inside the fore head move at a slower velocity than the continuous phase in the same radial position. As a result, the dispersed droplets impede the continuous phase surrounding the fore head of the droplets. On the other hand, the rear end will be dragged forward by the fore head due to the effect of interfacial tension, thus accelerating the continuous phase surrounding the rear end of the droplets. The accelerating effect of the rear end will be masked by the hindering effect of the fore head, which ultimately causes increased drag.

\begin{figure}
	\centering
	\includegraphics[width=0.45\linewidth]{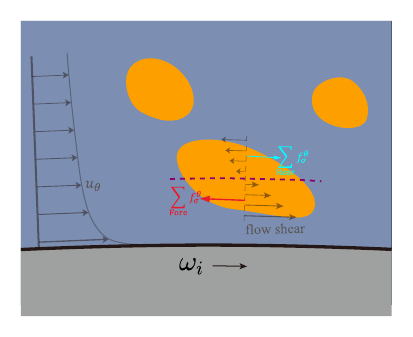}
	\caption{A sketch of how the interfacial tension works in a TC system.}
	\label{fig:51}
\end{figure}

To demonstrate our argument, we examine the interfacial tension experienced by droplets whose centers of mass fall within the range  $y^+<24$. This choice ensures that the majority of the droplets analyzed are in the viscous sublayer and buffer layer. 
For each individual droplet, we compute the sum of the azimuthal interfacial tension acting on the fore head, and plot the probability density function of these summed values for different droplets in figure \ref{fig:5}(\textit{b}). Similarly, we perform the analysis to the rear end of each individual droplet. In figure \ref{fig:5}(\textit{b}), the positive values indicate that the effect of interfacial tension appears to act in the flow direction, while the negative values suggest that the effect of interfacial tension appears to act against the flow direction. 
It can be observed that, for the fore head of the droplet, the effect of interfacial tension predominantly acts against the flow direction, which slows down the fluid inside the fore head, thus hindering the surrounding flow. Conversely, for the rear end, the effect of interfacial tension predominantly acts in the flow direction, which drives the fluid within the rear end to move faster, thus accelerating the surrounding flow.
To visualize the physical process in which the interfacial tension works, a sketch is shown in figure \ref{fig:51}.
Here, the effect of interfacial tension acting against the flow direction overwhelms that acting in the flow direction. 
The droplet-induced drag enhancement is dominated by the effect of interfacial tension on the fore head of the droplet, which leads to a hindering effect of the droplet on the surrounding continuous phase.

\begin{figure}
	\centering
	\includegraphics[width=1\linewidth]{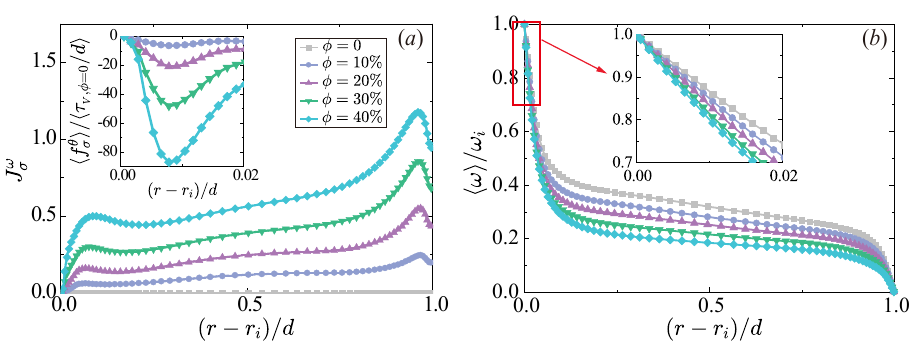}
	\caption{(\textit{a}) Interfacial tension contribution $J_\sigma^\omega(r)$ as a function of radial position at various droplet volume fractions. The inset shows the azimuthal component of the interfacial tension in the near wall region and the negative value indicates that the effect of interfacial tension acts against the flow direction. (\textit{b}) Average angular velocity $\left \langle \omega \right \rangle / \omega_i$ as a function of radial position at various droplet volume fractions with the inset showing the zoom-in of $\left \langle \omega \right \rangle / \omega_i$ near the IC.}
	\label{fig:52}
\end{figure}

Having gained an insight into how interfacial tension works, we are now able to comprehend the role of interfacial tension contribution $J^\omega_\sigma(r)$ to the total angular velocity flux, which is plotted in figure \ref{fig:52}(\textit{a}) for various volume fractions. Although the above discussion focuses on the region near the IC, the analysis of how interfacial tension works can be extended to the entire TC system. 
Note that in this paper the fore head always refers to the part of the droplet closer to the IC.
As shown in the inset of figure \ref{fig:52}(\textit{a}), the interfacial tension is negative in the region of $(r-r_{i})/d<0.02$, i.e., the effect of interfacial tension acting against the flow direction dominates in this region as just discussed. Specifically, the effect of interfacial tension results in a reduction in average angular velocity with increasing volume fraction in the region of $(r-r_{i})/d<0.02$ (see the inset of figure \ref{fig:52}\textit{b}). This leads to an increase in viscous stress with increasing volume fraction as shown in the inset of figure \ref{fig:2}(\textit{b}).
According to $J_{\sigma}^\omega(r) =-\int_{r_i}^{r} \left \langle {r^2 f_\sigma^\theta}\right \rangle  \, {\rm d}r$, $ J^\omega_\sigma(r)$ will increase until $\left \langle f^\theta_\sigma\right\rangle$ changes its sign from negative to positive, where $J^\omega_\sigma(r)$ attains its local maximum. After the peak, $J^\omega_\sigma(r)$ exhibits a slight decrease. The decrease of $J^\omega_\sigma(r)$ indicates that $\left \langle f^\theta_\sigma\right\rangle>0$. Near the IC, the droplets are highly deformed. When they are advected away from the IC, the droplets will relax back to a less deformed state. 
These cylindrical surfaces are mainly filled with the rear end of the highly deformed droplets and the fore head of the less deformed droplets. The slight decrease in $J^\omega_\sigma(r)$ can be attributed to the fact that the effect of interfacial tension experienced by the rear end of the highly deformed droplets is stronger compared to the that of the interfacial tension experienced by the fore head of the less deformed droplets. 

In the bulk region,  $J^\omega_\sigma(r)$ increases slightly with the radial position in the region of $0.2\lesssim (r-r_{i})/d\lesssim 0.8$, indicating that the total effect of the interfacial tension in that region is to impede the local flow. Near the OC, droplets are highly deformed due to the high shear near the solid surface. In this case the rear end of the droplet is closer to the OC where the angular velocity is zero, thus dragging the fore head backward. This leads to $\left \langle f^\theta_\sigma\right\rangle<0$ and $J^\omega_\sigma(r)$ obviously increases with the radial position in the region of $0.8\lesssim (r-r_{i})/d\lesssim 0.96$. On the other hand, the rear end will be dragged forward by the fore head, thus resulting in $\left \langle f^\theta_\sigma\right\rangle>0$ and $J^\omega_\sigma(r)$ shows an obvious decrease with the radial position in the region of $0.96\lesssim (r-r_{i})/d < 1$. At the OC with $(r-r_{i})/d = 1$, $J^\omega_\sigma(r)>0$, which indicates that the effect of interfacial tension acting against the flow direction dominates the modulation of the flow field within the cylinder gap. As a result, the overall reduction of the average angular velocity with increasing volume fraction is observed within the cylinder gap as shown in figure \ref{fig:52}\textit({b}).

\begin{figure}
	\centering
	\includegraphics[width=0.95\linewidth]{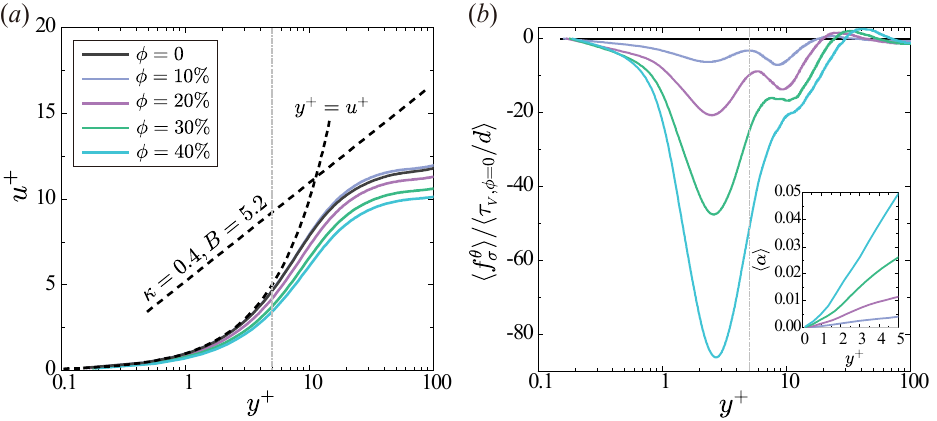}
	\caption{(\textit{a}) Mean azimuthal velocity profiles near the IC. $u^+=(u_i-\langle u_{\theta} \rangle)/u_\tau$ is the velocity difference from the IC normalized by the friction velocity. The dashed lines show the linear relation $u^+=y^+$ and the logarithmic law $u^+=(1/\kappa)\mathrm{ln}y^++B$ with the typical values $\kappa=0.4$ and $B=5.2$ ~\citep{huisman2013logarithmic}. (\textit{b}) Profiles for azimuthal component of the interfacial tension $\left\langle f^\theta_\sigma \right\rangle$. Inset: average phase fraction within the viscous sublayer. The vertical dotted line indicates the location of $y^+=5$. 
	}
	\label{fig:6}
\end{figure}

Based on the above analysis, we have attributed the drag enhancement to the effect of interfacial tension experienced by the fore head of the droplet, which acts against the flow direction and leads to a hindering effect of the droplet on the surrounding flow.
Although the division of the fore head and rear end based on the location of the droplet's center of mass is somewhat idealized, it is sufficient to reveal the physical mechanism of the droplet-induced drag enhancement.
The hindering effect is also evident in the mean azimuthal velocity. In figure \ref{fig:6}(\textit{a}), we present normalized mean azimuthal velocity profiles, \( u^+ = (u_i - \langle u_{\theta} \rangle) / u_\tau \), versus wall distance \( y^+ \).
In the single-phase case (\(\phi=0\)), \( u^+ \) follows the linear relation \( u^+ = y^+ \) effectively within the viscous sublayer (\( y^+ < 5 \)), indicating sufficient spatial resolution to resolve the boundary layer. At \( y^+ > 30 \), \( u^+ \) does not exhibit a clear logarithmic shape due to the low Reynolds number (\( Re \)) in this study \citep{huisman2013logarithmic}.
In the two-phase cases, \( u^+ \) shifts downward when \( \phi > 10\% \). This shift is observed not only in the buffer layer and above but also in the viscous sublayer, akin to drag enhancement observed with rough walls \citep{zhu2016direct, xu2023direct}. In the viscous sublayer, the magnitude of interfacial tension \( \left\langle f^\theta_\sigma \right\rangle \) increases with increasing volume fraction (figure \ref{fig:6}\textit{b} and its inset), rendering \( u^+ = y^+ \) invalid and indicating the significance of both viscous stress and interfacial tension. 
By examining the phase fraction distribution, we find that the droplet phase fraction can reach a maximum of $5\%$ inside the viscous sublayer. In addition, the magnitude of the azimuthal interfacial tension reaches its maximum at about $y^+=3$, indicating that the azimuthal interfacial tension has the greatest effect inside the viscous sublayer. This finding is consistent with previous experimental results of \cite{yi2022physical}. The authors found that the droplets are fragmented within the boundary layer. Hence, the feedback of droplets will attain its maximum within the boundary layer. The downward shift of $u^+$ may be due to the increased effective viscosity by interfacial tension, causing $y^+$ to be overestimated. Future theoretical studies on turbulent boundary layers in two-phase flows should consider the impact of interfacial tension in the viscous sublayer. 

Our findings emphasize the dominant role of interfacial tension in the drag enhancement caused by droplets. In industrial applications, many existing models (e.g., Euler-Lagrange approach) may not adequately account for the effects of interfacial tension. Our results show that to accurately predict drag and flow characteristics in two-phase turbulent flow systems, models must account for the contribution of interfacial tension, particularly in the boundary layer region. This provides guidance for the improvement and development of two-phase flow models.

The budget analysis adopted in this work is widely used to analyze the drag modulation in two-phase turbulence. In turbulent channel flow laden with dense suspensions ($\phi=0-20\%$) of neutrally buoyant spheres, ~\cite{picano2015turbulent} demonstrated that the drag enhancement is dominated by the particle-induced stress contribution. Interestingly, similar to the profile of $J^\omega_\sigma(r)$ near the IC, the particle-induced stress contribution is positive and increases with wall distance, then decreases further away from the wall~\citep{picano2015turbulent,wang2017modulation}. In turbulent flow laden with particles, the exact expression for the particle-induced stress remains unknown. Inspired by the interfacial tension contribution induced by the droplets, which can be considered as “deformable particles”, we can replace the particle-induced stress effect with the integral of the particle-induced force in the direction normal to the wall. This means that, in the near wall region, the particle-induced stress acts against the flow direction, thus hindering the surrounding continuous phase. Based on the above analysis, we infer that the drag enhancement induced by droplets and particles is dominated by similar mechanisms within the boundary layer. 
Our results provide one possible explanation for why neutrally buoyant solid particles have a stronger drag enhancement effect compared to droplets as found by~\cite{de2019effect} and~\cite{yi2021global}. The stronger drag enhancement effect indicates that the hindering effect on the surrounding flow field near the wall is more pronounced for solid particles, which causes larger viscous stress. This may be due to the fact that the deformability of the droplet allows them to adjust their shape when subject to shear flow, thus droplets affect the surrounding flow to a less degree. 
As a result, the drag enhancement caused by droplets is less pronounced than that caused by solid particles. 

Given that our work is primarily aimed at reproducing the drag enhancement reported in our previous experimental study and revealing the underlying mechanism, the dimensionless parameters in our simulations mainly correspond to $Re=5.2 \times 10^3$ cases in experiments \citep{yi2021global}. Although we kept many fixed dimensionless parameters ($Re, Ta, \Gamma, \eta, We$, density ratio, viscosity ratio, contact angle) in our simulations, we can propose some new explanations for the effects of different parameters based on our findings.
By changing the angular velocity of the inner cylinder, it is observed that the effective viscosity decreases with increasing $Re$ ($Ta$ or $We$), suggesting a shear thinning effect \citep{yi2021global}. This can be attributed to the fact that the effect of interfacial tension weakens as the turbulence intensity increases. Specifically, the increase in the near-wall viscous stress due to interfacial tension gradually diminishes relative to the increase in viscous stress due to the increase in turbulence intensity. The effects of the density ratio and viscosity ratio have been studied in our previous work~\citep{su2024numerical}. 
We found that decreasing the density and viscosity ratios of the dispersed phase to the continuous phase reduces the contribution of local advection and diffusion terms to the momentum transport, respectively, resulting in drag reduction. The change in contact angle may cause droplets to attach to the inner cylinder and form a lubricating layer, thus contributing to the drag reduction. Herein, the effect of the attached droplets on the system drag is dominated by the interfacial tension on the rear end, which appears to act in the flow direction. 
Specifically, the effect of interfacial tension tends to increase the angular velocity of the fluid near the interface of the attached droplets and decrease near-wall viscous stress, thus contributing to drag reduction. Therefore, the droplet phase attached to the inner cylinder and dispersed within the system will compete to cause drag modulation. For the parameters characterizing the TC device ($\eta$ and $\Gamma$), the choice of aspect ratio of $\Gamma=2\pi/3$ is to minimize computational cost and corresponds to the cases where the axial length is much larger than the gap width in real experiments. The choice of curvature ratio $\eta=r_i/r_o=0.714$ may limit our findings to only TC systems with curvature ratios close to this value. Further studies are needed to extend the mechanisms of interface-induced drag modulation to TC systems with other curvature ratios and even to channel flow systems.

\section{Conclusions}

In this study, we investigated the mechanism of drag enhancement by neutrally buoyant droplets in liquid-liquid Taylor--Couette turbulence at a Reynolds number of 5200 and a Weber number of around 8. Our focus was on the effect of droplet presence within the boundary layer at moderate volume fractions. To achieve this, we employed a modified multi-marker Volume-of-Fluid (VOF) method, which accurately reproduces the drag enhancement effect and aligns well with experimental results.

Through angular velocity flux analysis, we identified that interfacial tension plays a pivotal role in drag enhancement. To understand how interfacial tension influences drag enhancement, we examined the effects of droplets on viscous stress and angular velocity near the inner cylinder. Our observations revealed that droplets experience significant deformation and stretching along the streamwise direction due to shear near the inner cylinder. Consequently, the rear end of the droplets lags behind the fore head. To keep the integrity of the droplets and resist deformation, the effect of interfacial tension experienced by the fore head of the droplets appears to act against the flow direction while the effect of interfacial tension experienced by the rear end appears to act in the flow direction.
The effect of the interfacial tension acting against the flow direction overwhelms that of the interfacial tension acting in the flow direction, leading to a hindering effect of the droplets on the surrounding continuous phase. This hindering effect alters high-speed streaks, reducing their size and increasing the occurrence of low-speed streaks, which typically contribute to higher viscous stress and system's drag.
Furthermore, we observed that the mean streamwise velocity profile no longer follows the linear relation $u^+ = y^+$ in two-phase turbulence, which may be due to the fact that the interfacial tension increases the effective viscosity, resulting in an overestimation of $y^+$.
Our findings underscore the necessity of adequately considering interfacial tension in the near-wall region when modeling two-phase turbulence.

\backsection[Acknowledgements]{This work is financially supported by the National Natural Science Foundation of China under Grant Nos. 11988102, 22478421, 12402299 and 12402298, the New Cornerstone Science Foundation through the New Cornerstone Investigator Program and the XPLORER PRIZE, and the Science Foundation of China University of Petroleum, Beijing (No. 2462024YJRC008). We would like to thank Dr. Feng Wang, Dr. Junyi Li and Dr. Mingbo Li for their valuable suggestions.}

\backsection[Declaration of Interests]{The authors report no conflict of interest.}

\backsection[Author ORCIDs]{\\
Jinghong Su https://orcid.org/0000-0003-1104-6015;\\
Yi-Bao Zhang https://orcid.org/0000-0002-4819-0558; \\
Cheng Wang https://orcid.org/0000-0002-6470-7289;\\
Lei Yi https://orcid.org/0000-0002-0247-4600;\\
Fan Xu https://orcid.org/0009-0004-3324-7859;\\
Yaning Fan https://orcid.org/0009-0000-5886-3544; \\
Junwu Wang https://orcid.org/0000-0003-3988-1477; \\
Chao Sun https://orcid.org/0000-0002-0930-6343.}

\appendix
\section{Resolution test}\label{appA}
\begin{figure}
	\centering
	\includegraphics[width=0.6\linewidth]{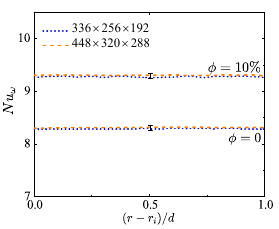}
	\caption{Radial dependence of $Nu_\omega$ for two different grid resolutions. An error bar indicating a $1\%$ error is provided for reference and both the cases for $\phi=0$ and $\phi=10\%$ lie within the error bar.
	}
	\label{fig:S01}
\end{figure}

To obtain reliable numerical results, the grid's spatial resolutions have to be sufficient. The requirement for spatial resolution is to have the grid spacing in each direction of the order of Kolmogorov length $\eta_k$. For the single-phase flow with $\phi=0$, a reasonably resolved case ($N_\theta \times N_r \times N_z = 336\times256\times192$) with the maximum grid spacing smaller than 1.39$\eta_k$ and an extremely well-resolved case ($N_\theta \times N_r \times N_z = 448\times320\times288$) with the maximum grid spacing smaller than $\eta_k$ are considered for resolution test as depicted in figure~\ref{fig:S01}. The same resolution test is also conducted for two-phase flow with $\phi=10\%$.
Both the cases for $\phi=0$ and $\phi=10\%$ lie within $1\%$ error bar, indicating that the adopted spatial resolution $N_\theta \times N_r \times N_z = 336\times256\times192$ is sufficient to obtain reliable results. 

\section{Feasibility analysis of the modified multi-marker VOF method}\label{appB}
As one of the ways to avoid numerical coalescence in interface-capturing methods, the multi-marker method has been effectively used in liquid-liquid and gas-liquid two-phase turbulence simulations~\citep{balcazar2015multiple,hasslberger2020direct,nemati2021direct}. In brief, the method assigns a separate phase fraction to each bubble or droplet to avoid numerical coalescence, which would often occur for the single-marker formulation in the standard VOF method. However, due to the greatly increased computational cost of the multi-marker method, it is difficult to simulate cases with droplet numbers of hundreds or more, leading to difficulties in simulating moderate or high droplet volume fractions. 

 \begin{figure}
	\centering
	\includegraphics[width=0.6\linewidth]{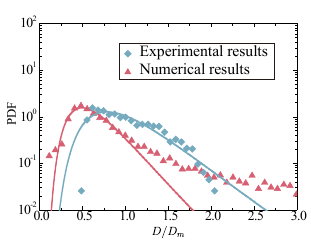}
	\caption{The PDF of the droplet diameter $D$ with respect to the average diameter $D_m$ for two-phase turbulence with $\phi = 40\%$. The solid lines denote the ﬁtting results with a log-normal distribution function. The simulated results are compared with experimental results from our previous work~\citep{yi2021global}. 
	}
	\label{fig:S02}
\end{figure}
 \begin{figure}
	\centering
	\includegraphics[width=1\linewidth]{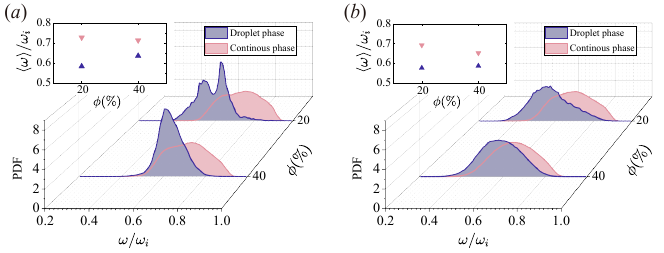}
	\caption{Angular velocity probability density functions for the droplet and continuous phases at the radius cut $(r_{cut}-r_i)/d =0.02$ obtained using (\textit{a}) the standard VOF method and (\textit{b}) the modified multi-marker VOF method. The inset shows the average angular velocity of the droplet and continuous phases. 
	}
	\label{fig:S03}
\end{figure}

In this work, we have modified the multi-marker method by replacing the assignment of a separate phase fraction to each droplet with a separate phase fraction for each group of droplets, thus allowing the simulation of two-phase flow at moderate droplet volume fractions. 
From our estimates, simulation of two-phase flow with volume fractions of $\phi=10\%$, $20\%$, $30\%$, and $40\%$ using the modified multi-marker VOF method requires nearly 1.43, 2.39, 3.27, and 4.21 times that of the standard VOF method, respectively. For simulations with $\phi=10\%$, $20\%$, $30\%$ and $40\%$, each time step takes nearly 0.19 core hours, 0.32 core hours, 0.44 core hours and 0.56 core hours, respectively.
Unfortunately, the modified multi-marker VOF method does not completely resolve the issue of numerical coalescence, as shown in figure~\ref{fig:S02}. Due to numerical coalescence, the simulated probability density distribution of droplet sizes deviates from the experimental results. Therefore, further optimization of the simulation method is needed in future work to accurately simulate the probability density distribution of droplets at moderate volume fractions. Despite the influence of numerical coalescence on the simulated drag enhancement, particularly at volume fractions of $30\%$ and $40\%$ as shown in figure~\ref{fig:1}(\textit{a}), we have successfully captured the droplet-induced drag enhancement within an acceptable range of error. Our codes are available at https://github.com/Sujh123/Multi-marker-VOF-method.

To ensure that the method does not qualitatively change the mechanism of droplet-induced drag enhancement, we additionally calculate the droplet volume fractions of $20\%$ and $40\%$ using the standard VOF method, as shown in figure~\ref{fig:S03}(\textit{a}). 
These results are qualitatively consistent with those obtained using the modified multi-marker VOF method as shown in figure~\ref{fig:S03}(\textit{b}), i.e., the average angular velocities of the dispersed phase increases with the droplet volume fraction while the average angular velocities of the continuous phase decreases with the droplet volume fraction.
In addition, it is observed that the average angular velocities of both the dispersed and continuous phases within the radius cut obtained by the standard VOF method are larger than those obtained by the multi-marker VOF method, especially at a droplet volume fraction of $40\%$, which is attributed to the larger droplet sizes due to the effect of numerical unphysical coalescence in the standard VOF method.  
This leads to a weaker ability to resist droplet deformation, reducing the ability to impede the flow of the surrounding continuous phase. These results indirectly support the conclusions drawn from this work.

\section{Spurious current}\label{appC}

\begin{figure}
	\centering
	\includegraphics[width=0.95\linewidth]{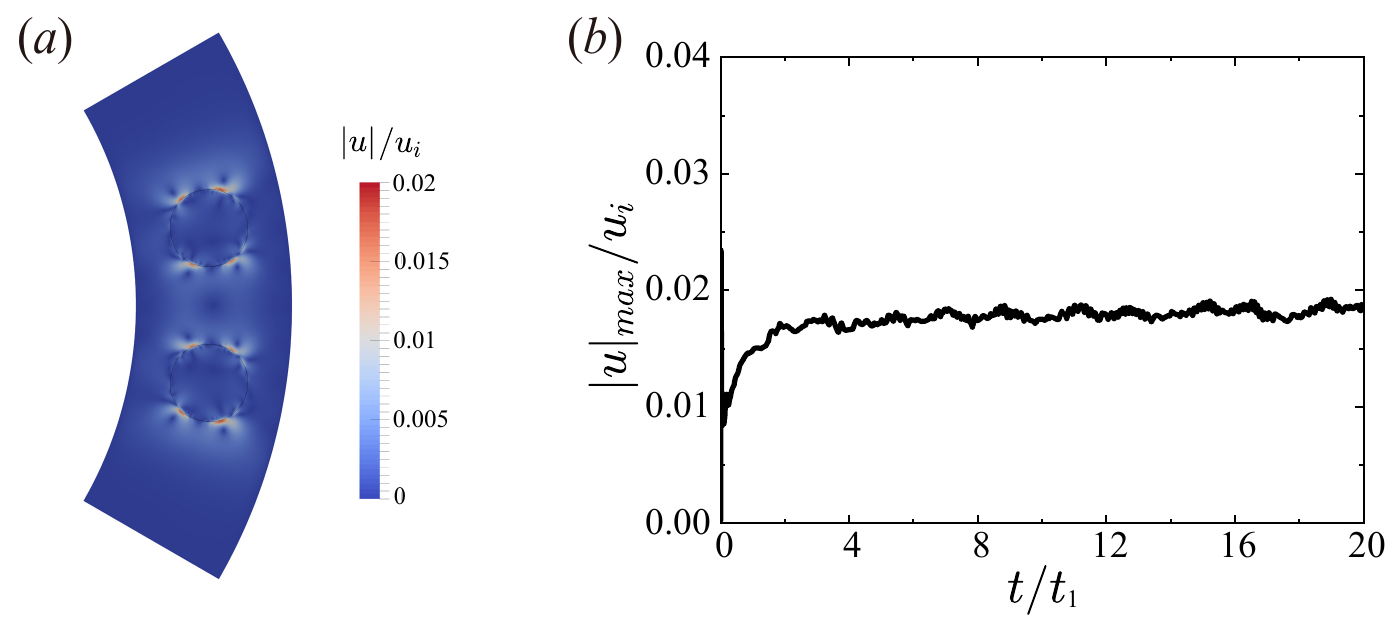}
	\caption{Spurious currents for two static drops in a TC system with the two cylinders fixed. (\textit{a}) Contour of the velocity magnitude. (\textit{b}) The maximum velocity magnitude, $|u|_{max}$, as a function of time. $|u|_{max}$ is normalized by the velocity of the inner cylinder $u_i$ considered in our work and the time $t$ is normalized by the time required for one full rotation of the inner cylinder $t_1=2\pi r_i/u_i$.
	}
	\label{fig:S04}
\end{figure}

In interface-capturing methods, it is usually non-trivial to compute the interface curvature, thus making it difficult to achieve a strict Laplace balance between surface tension and pressure gradient. This leads to the production of numerical artifacts called spurious or parasitic currents. To assess the effect of the numerical artefacts when using the modified multi-marker VOF method, we show the velocity field induced by the spurious currents in a two-dimensional cylindrical domain (see figure~\ref {fig:S04}). Figure~\ref{fig:S04}(\textit{a}) shows a snapshot of two static drops with different markers. 
The inner and outer cylinders are fixed and the two drops have a diameter being half of the gap width. It is observed that the spurious currents mainly appear near the two-phase interface. After normalizing the maximum velocity magnitude using the velocity of the inner cylinder adopted in our work, the $|u|_{max}/u_i$ is below 0.02. This is acceptable in our simulations since the simulated drag enhancement is consistent with the experimental results and the radial dependence of $Nu_\omega$ is less than $1\%$ across the cylinder gap as shown in figure 9, which is a very stringent requirement for numerical convergence in TC system. 
Although there are several methods to evaluate surface curvature more accurately \citep{soligo2021turbulent}, it is difficult to satisfy both balance and momentum conservation requirements \citep{popinet2018numerical}.
In addition, the cylindrical system we studied poses a greater challenge to existing methods.
Considering that our system needs to strictly guarantee momentum conservation and that the effect of the spurious currents on the system's drag can be neglected, the simple continuum surface force model is adopted in our work.

\bibliographystyle{jfm}
\bibliography{jfm-instructions}

\end{document}